\documentclass{revtex4}

\usepackage{graphicx}
\newcommand {\gv} {\tilde{\mathrm{G}}}
\def\md{M_*^{(n)}}
\def\gev{\, {\rm GeV}}

\def\beq{\begin{equation}}
\def\eeq{\end{equation}}
\def\bea{\begin{eqnarray}}
\def\eea{\end{eqnarray}}
\def\Emiss{\not  \! \! E}

\def\leqn#1{(\ref{#1})}

\def\stacksymbols #1#2#3#4{\def\theguybelow{#2}
    \def\vp{\lower#3pt}
    \def\sp{\baselineskip0pt\lineskip#4pt}
    \mathrel{\mathpalette\intermediary#1}}

\def\intermediary#1#2{\vp\vbox{\sp
     \everycr={}\tabskip0pt
     \halign{$\mathsurround0pt#1\hfil##\hfil$\crcr#2\crcr
              \theguybelow\crcr}}}

\def\lapproxeq{\stacksymbols{<}{\sim}{2.5}{.2}}

\begin{document}
% You should use BibTeX and revtex.bst for references
\bibliographystyle{revtex}

% number  and your conference paper identification number on the
% title page in preprint mode. Multiple \preprint commands are allowed.

\begin{flushright}
hep-ph//0110339 \\
LBNL-49069 \\
\end{flushright}

\title{Extra dimensions vs.\ supersymmetric interpretation \\
of missing energy events at a linear collider}
% Optional argument for running titles on pages
%\title[]{}



\author{Shrihari Gopalakrishna$^{a,b}$, Maxim Perelstein$^{b}$, James D. Wells$^{a,b}$}
%\thanks{}
\affiliation{
${}^{(a)}$Physics Department, Univ.\ of California, Davis, CA 95616 \\
${}^{(b)}$Theory Group, Lawrence Berkeley National Lab, Berkeley, CA 94720 
}


\date{\today}

\begin{abstract}
The photon plus missing energy signature is a primary handle on two
important classes of theories. Theories with large extra dimensions
predict the production of photons in association with Kaluza-Klein excitations
of the graviton. In supersymmetric theories with superlight gravitinos, 
photons can be produced in association with gravitino pairs.  The signatures 
for these two theories are compared, and it is found that they can be
distinguished by studying the photon energy distributions and scaling 
of the cross section with center-of-mass energy. Both these methods fail,
however, if there are six extra dimensions. 
In that case, additional phenomena predicted by the theories would be
required to narrow down the underlying causes of the photon plus missing
energy signal. We also study the ability of these measurements to 
determine the number of extra dimensions.

\end{abstract}
% \pacs{}

%\maketitle must follow title, authors, abstract and \pacs
\maketitle

\section{Introduction}
%\label{}

%\subsection{}
%\subsubsection{}

In this report we assume that a signature for new physics is observed at 
a linear collider in the photon + missing energy channel,
\beq
e^+e^- \rightarrow \gamma + \Emiss.
\label{p311signal}
\eeq 
We analyze two 
possible interpretations of this signal. One possibility is in the 
context of the proposal of Arkani-Hamed, Dimopoulos and
Dvali (ADD)~\cite{Arkani-Hamed:1998rs} that we live in a world 
with extra spatial 
dimensions. In this proposal, the gauge hierarchy problem is resolved by 
assuming that the fundamental, higher-dimensional gravitational scale is 
around a TeV. The apparent high value of the Planck scale in the effective  
four-dimensional theory can be explained if the volume of the extra dimensions
is large. The Standard Model (SM) degrees of freedom are confined to a 
four-dimensional manifold (a ``brane'') in the full space-time, while 
gravitons can propagate in the extra dimensions. If an extra-dimensional 
graviton is emitted in a particle collision, it would not be
detected, leading to the signature of~\leqn{p311signal}.    

Another possibility is in the context of an effective supersymmetric theory 
obtained from a spontaneously broken supergravity theory. If the scale of 
supersymmetry breaking is sufficiently low such theories will contain an 
extremely light gravitino, 
$ m_{\tilde{G}}\sim10^{-3}$~eV. Gravitino pair production in 
conjunction with a photon will again produce the signature 
of~\leqn{p311signal}.    

In this work we ask how well we can interpret a photon + missing energy 
signal. In particular, we would like to know whether one can distinguish 
between the two possibilities discussed above. We will also study if it
is possible to distinguish between the ADD-type models with various numbers 
of extra dimensions.

%%%%%%%%%%%%%%%%%%%%%%%%%%%%%%%%%%%%%%%%%%%%%%%%%%%%%%%%%%%%%%%%%%%%%%
\section{Gravitons in Extra dimensions}

In theories with large extra dimensions,
the missing energy in $e^+e^- \rightarrow \gamma + \Emiss$ could be carried 
by the Kaluza-Klein(KK) excitations of the gravitons. The differential 
cross-section is given by~\cite{Giudice:1999ck}\cite{Mirabelli:1999rt}:
\beq
\frac{d^2\sigma}{dx_\gamma d\cos \theta}(e^+e^- \to \gamma G)
=\frac{\alpha}{32}~S_{n -1}~ 
\left( \frac{\sqrt{s}}{\md}\right)^{n +2}
\frac{1}{s}f(x_\gamma, \cos\theta)
\label{p311sezgg}
\eeq
\beq
f(x,y)=\frac{2(1-x)^{\frac{n}{2} -1}}{x(1-y^2)}
\left[ (2-x)^2(1-x+x^2)-3y^2x^2(1-x)-y^4x^4 \right] .
\label{p311funz}
\eeq
Here $M_*^{(n)}$ is the fundamental mass scale, $n$ is the number of 
extra dimensions, $S_{n -1 }$ is the surface area of an 
{$n$}-dimensional sphere of unit radius, 
$x_\gamma =2E_\gamma /\sqrt{s}$, $E_\gamma$ is the photon energy, 
and $\theta$ is the angle between the photon and beam directions.

%%%%%%%%%%%%%%%%%%%%%%%%%%%%%%%%%%%%%%%%%%%%%%%%%%%%%%%%%%%%%%%%%%%%%%%%%
\section{Gravitino signal}

In supersymmetric theories with a superlight gravitino, photon + missing 
energy events could arise due to the pair-production of gravitinos, 
$e^+e^- \rightarrow \gv\gv\gamma$. The differential cross-section is given 
by~\cite{Brignole:1998sk}:
\beq
  \frac{d^2\sigma}{dx_{\gamma}\,d\cos\theta} = 
  \left( \frac{\alpha G_N^2}{45} \right) 
  \frac{s^3}{m_{\gv}^4} f_{\gv\gv\gamma}(x_{\gamma},\cos\theta)
  \label{p311eq:GGg1}
\eeq
\beq
  f_{\gv\gv\gamma}(x,\cos\theta) = 
        2(1-x)^2 \left[ \frac{(1-x)(2-2x+x^2)}{x\sin^2\theta} +
                \frac{x(-6+6x+x^2)}{16} - \frac{x^3\sin^2\theta}{32} \right]
  \label{p311eq:GGg}
\eeq
where $\alpha$ is the fine structure constant, $G_N$ is the 
gravitational constant, $m_{\gv}$ is the gravitino mass, 
$x_{\gamma}$~is the photon scaled energy ($E_\gamma/E_{\rm beam}$) 
and $\theta$ is the polar angle.

%%%%%%%%%%%%%%%%%%%%%%%%%%%%%%%%%%%%%%%%%%%%%%%%%%%%%%%%%%%%555
\section{Observations}

Single photon plus missing energy events also occur within the
SM due to escaping neutrinos.  The SM background
is $e^+e^- \rightarrow \nu \bar\nu \gamma$, and
can be obtained from numerous Monte Carlo packages, including 
Pandora~\cite{Peskin:1999hh}.
Previous analyses~\cite{Giudice:1999ck,Mirabelli:1999rt,Brignole:1998sk} 
have carefully outlined the parameter spaces that
produce a discernible excess of signal events over SM background in
extra dimensional gravity models and in superlight gravitino models.
We do not reproduce those results here, but are more interested in
how similar or dissimilar the graviton-induced signal is to the
gravitinos-induced signal.  

\begin{figure}
\includegraphics[scale=0.6]{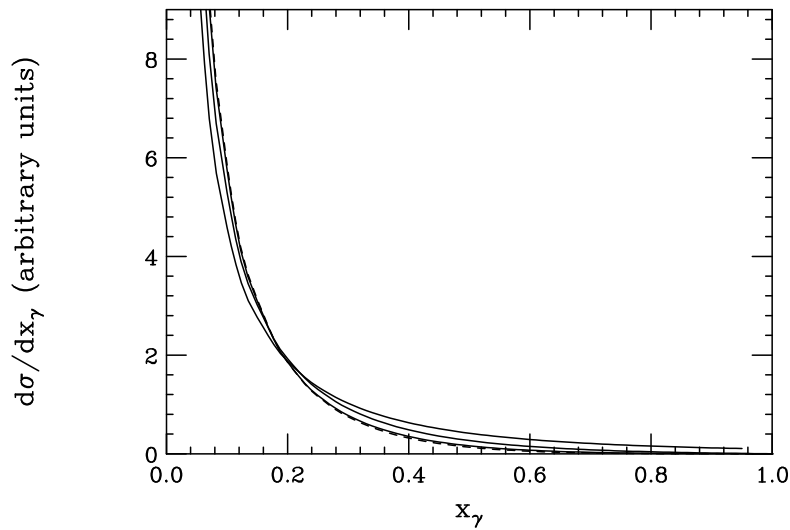}%
\caption{The differential cross-section as a function of the fractional photon
missing energy.  The scales $M_*^{(n)}$ and $m_{\gv}$ are
tuned so that the total cross-section is the same for each model.
The three solid lines are for the graviton KK signal with 2, 4 and 6
extra dimensions.  At $x_\gamma=0.5$ the ordering of these solid lines
is  $n=2$ highest, and $n=6$ lowest.  The dashed line is 
the gravitino signal.}
\label{p311diffcs1}
\end{figure}

\begin{figure}
\includegraphics[scale=0.6]{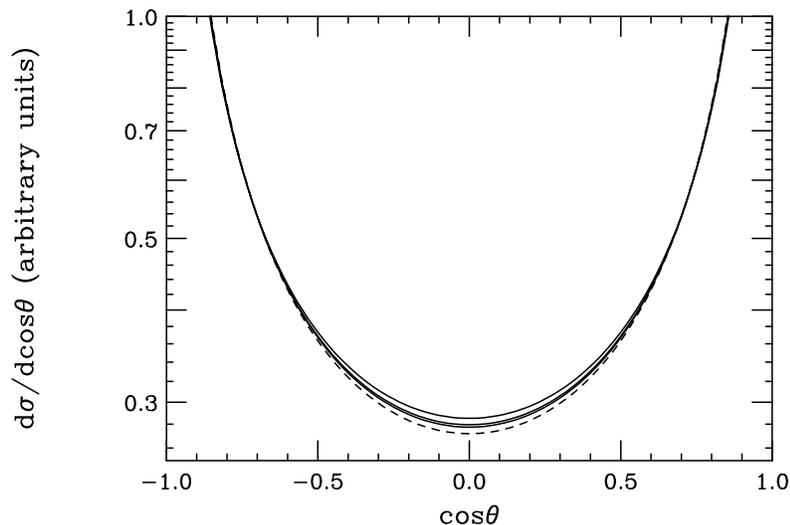}%
\caption{The differential cross-section as a function of the 
photon $\cos\theta$. The scales $M_*^{(n)}$ and $m_{\gv}$ are
tuned so that the total cross-section is the same for each model. 
The three solid lines are for the graviton KK signal with 2, 4 and 6
extra dimensions.  At $\cos\theta=0$ the ordering of these solid lines
is  $n=2$ highest, and $n=6$ lowest.  The dashed line is 
the gravitino signal.}
\label{p311diffcs2}
\end{figure}

To help us see the difference between the possible signal interpretations
of single photon plus missing energy events at a linear collider, we have
plotted in Fig.~\ref{p311diffcs1} the differential cross-section as a 
function of the photon missing energy, $x_\gamma =2E_\gamma /\sqrt{s}$, 
obtained by integrating the cross-sections~\leqn{p311sezgg} 
and~\leqn{p311eq:GGg1} with respect to $\cos\theta$. 
The three solid lines are for the graviton KK signal with 2, 4 and 6
extra dimensions. At $x_\gamma=0.5$ the ordering of the solid lines
is  $n=2$ highest, and $n=6$ lowest. The dashed line is the gravitino signal.
The scales ($M_*^{(n)}$ for the ADD-type models and
$m_{\gv}$ for the supersymmetric model) have been chosen so that the
total cross sections within the applied kinematic cuts, $x_\gamma>
0.05$ and $|\cos \theta|<0.95$, are equal in all four cases. 
Similarly, Fig.~\ref{p311diffcs2} shows the normalized 
differential cross-sections of the four models as a function of $\cos\theta$, 
obtained by integrating with respect to $x_\gamma$. The ordering of the solid 
lines at $\cos\theta=0$ is $n=2$ highest, and $n=6$ lowest.  

%-------------------------------------------------------------------------

%------------------------------------------------------------------------- 

%%
It is clear from the figures that both the recoil energy and
angular distributions of the events are quite similar in all
four cases. It is still possible, however, to use the photon energy 
distribution to discriminate between the ADD models with different values 
of $n$. To do this, we perform a harder cut on the photon energy,
$x_\gamma > 0.2$. This is advantageous because in this region, the 
differences between cross 
sections with different $n$ do not change sign. We also perform 
an additional cut on $x_\gamma$ to get rid of the peak in the
Standard Model background associated with on-shell $Z$ production. 
Assuming that electron and positron beam polarizations of 
80 and 60\% respectively are achieved at a 500 GeV linear collider, and 
that the
cross section can be measured with 1\% precision, we find that,  
for example, the case $n=2$ can be distinguished from $n=3$ for
values of $M_*^{(2)}$ up to 4.6 TeV. (If the uncertainty is purely
statistical, 1\% error on the cross section measurement corresponds to 
integrated luminosity of about 270 fb$^{-1}$.) 
Note, however, that the photon energy distributions for the cases
of gravitino emission and $n=6$ ADD model are extremely close. 
It is not likely that the ambiguity between these two cases can be 
resolved by this measurement. 

We have also studied whether performing a more restrictive angular cut
could help in distinguishing between the cases of gravitino and $n=6$ 
graviton emission. From Fig.~\ref{p311diffcs2} it appears that the two
signals have discernible differences at small values of $\cos\theta$.        
We have found, however, that with the assumed 1\% precision on the 
cross section measurement, this analysis will not be able to 
conclusively separate these two cases. 

Another way to distinguish between models is by varying the collider 
center of mass energy $\sqrt{s}$. The TESLA study~\cite{TESLA} has 
shown that measuring cross sections at two design center of mass 
energies, $500\gev$ and $800\gev$, allows to determine the number of 
extra dimensions in the ADD models for a wide range of parameters. For
example, if there are two extra dimensions, the $n=3$ case can be 
excluded at 99\% CL for values of $M_*^{(2)}$ up to 6.1 TeV. The beam 
polarizations assumed in~\cite{TESLA} coincide with the values chosen
for our analyses here. 

While upgrading the linear collider energy from 500 GeV
to 800 GeV, for example, will take much time, effort and money, the beam
energy can be changed by as much as 5\% without significantly disrupting
the operation of the collider. Therefore, it is interesting to study if 
different models can be distinguished by performing measurements at
two center of mass energies separated by 5\%. We have performed 
such a study using Pandora. We have chosen the scales of each model so that
their cross sections (with the same cuts and polarizations as before) are 
identical at $\sqrt{s}=500$ GeV, and evaluated their differences at  
$\sqrt{s}=475$ GeV. Taking the Standard Model background into account and 
assuming that the cross sections at each energy can be measured with
1\% precision, we find that the $n=2$ and $n=3$ cases can be distinguished 
at the 3$\sigma$ level for $M_*^{(2)} \lapproxeq 3.9$ TeV. The sensitivity of 
this method is thus somewhat lower than for studying photon energy 
distributions. Surprisingly, 
the scaling of the gravitino pair production cross section with $\sqrt{s}$ 
is again identical to the $n=6$ ADD model. Changing collider 
energy does not help in separating these two cases. 

We conclude that while the ADD models with $n \not= 6$ can be distinguished
from the supersymmetric models with light gravitino by carefully 
studying the photon + missing energy signal alone, this cannot be 
done for the case of 6 extra dimensions. (Let us note in passing that 
the missing energy signal in the $n=6$ ADD model is also identical
to the one due to the emission of scalar states associated with the 
brane coordinates~\cite{Creminelli:2001gh}.) In this case,  
distinctions between the models will only occur with additional
phenomena.  The next-best observables in the extra-dimensional graviton
model include the dimension-eight contact 
interactions~\cite{Giudice:1999ck,Hewett:1999sn} induced by 
\beq
e^+e^-\to G^{(n)}\to \gamma\gamma,\, f\bar f,\, {\rm etc.,}
\eeq
or from string Regge states~\cite{Dudas:2000gz,Cullen:2000ef}.

The superlight gravitino models are supersymmetric models, and the
superpartner spectrum must satisfy the same naturnalness criteria
as other supersymmetry ideas.  The expectation is that some superpartner
states, most notably charginos and neutralinos, 
should be accessible and well-studied
at a 500 GeV linear collider~\cite{Abe:2001wn}. The kinematic reach
for superpartners then approaches $\sqrt{s}/2=250\gev$ in $e^+e^-$
collisions.  An increase of the reach of selectrons is possible
in $e^-\gamma\to \tilde G\,\tilde e_{R,L}$ collisions, since selectron
masses approaching $\sqrt{s}=500\gev$ are kinematically 
accessible~\cite{Gopalakrishna:2001cm}.
The full complement of observables will be needed to distinguish
the precise underlying theory being discovered at the colliders.

\section{Acknoledgements}

We would like to thank the organizers of the Snowmass 2001 workshop
where this work was initiated. We are grateful to many of the workshop's
participants and especially to the members of the P3 and E3 working groups
for interesting discussions. We are indebted to Michael Peskin for his 
invaluable help with the Pandora event generator.

This work was supported in part by the Director, Office of Science,
Office of High Energy and Nuclear Physics, of the U.S. Department of
Energy under Contract DE-AC03-76SF00098. 

%%%%%%%%%%%%%%%%%%%%%%%%%%%%%%%%%%%%%%%%%%%%%%%%%%%%%%%%%%%%%%%%%%%%%%%%

\end{document}